\documentclass[9pt,superscriptaddress]{revtex4-2}

\setcitestyle{super,sort&compress,comma}
\usepackage{amsmath} 
\usepackage{amsfonts}
\usepackage{amssymb}
\usepackage{braket} 
\usepackage{graphicx} 
\usepackage{hyperref} 
\usepackage[noabbrev]{cleveref}
\usepackage{wrapfig} 
\usepackage[nottoc,numbib]{tocbibind}
\usepackage{booktabs}
\usepackage[utf8]{inputenc} 
\usepackage{float} 
\usepackage{charter}

\begin{document}
\title{From Goldilocks to Twin Peaks: multiple optimal regimes for quantum transport in disordered networks}
\author{Alexandre R. Coates}
\affiliation{SUPA, Institute of Photonics and Quantum Sciences, Heriot-Watt University, Edinburgh EH14 4AS, United Kingdom}
\author{Brendon W. Lovett}
\affiliation{SUPA, School of Physics and Astronomy, University of St Andrews, St Andrews KY16 9SS, United Kingdom} 
\author{Erik M. Gauger}
\email{e.gauger@hw.ac.uk}
\affiliation{SUPA, Institute of Photonics and Quantum Sciences, Heriot-Watt University, Edinburgh EH14 4AS, United Kingdom}

\begin{abstract}
Understanding energy transport in quantum systems is crucial for an understanding of light-harvesting in nature, and for the creation of new quantum technologies. Open quantum systems theory has been successfully applied to predict the existence of environmental noise-assisted quantum transport (ENAQT) as a widespread phenomenon occurring in biological and artificial systems. That work has been primarily focused on several `canonical' structures, from simple chains, rings and crystals of varying dimensions, to well-studied light-harvesting complexes. Studying those particular systems has produced specific assumptions about ENAQT, including the notion of a single, ideal, range of environmental coupling rates that improve energy transport. In this paper we show that a consistent subset of physically modelled transport networks can have at least two ENAQT peaks in their steady state transport efficiency.
\end{abstract}
\maketitle

\twocolumngrid

\section{Introduction}
\label{sec: intro}
Energy transport occurs in many contexts: from circuits and molecular junctions to processes like photosynthesis and the electron transport chain in biology \cite{Kundu2017NanoscaleHarvesting, Kassal2013DoesPhotosynthesis,Amarnath2016MultiscalePlants,Bennett2013ADescription,Chavez2021Disorder-EnhancedCavities,Brixner2017ExcitonSystems, Spano1991CooperativeAggregates, Saikin2013PhotonicsAggregates}. 
In 2008 the specifics of Environmental Noise-Assisted Quantum Transport (ENAQT) were first laid out in both artificial and natural schema~\cite{Plenio2008Dephasing-assistedBiomolecules, Caruso2009HighlyTransport, Cao2009OptimizationProcesses, Mohseni2008Environment-assistedTransfer}. Since then there has been a proliferation of research into the interplay of coherent quantum transport and noise in many systems~\cite{Huelga2013VibrationsBiology,Kassal2012Environment-assistedSystems, Caruso2014UniversallyNetworks}, such as the role environmental phonon coupling can have in overcoming the effects of coherent localisation~\cite{Coates2021LocalisationTransport, Chin2010Noise-assistedComplexes, Zerah-Harush2020EffectsTransport, Olaya-Castro2008EfficiencyCoherence}. From these studies and others, an intuition has been established that ENAQT produces a single `Goldilocks zone'~\cite{Mohseni2014Energy-scalesComplexes, Harris2017QuantumProcessor, Shabani2012EfficientComplexes} where dephasing overcomes limits inherent to fully coherent dynamics, but is not sufficiently aggressive to spoil its favourable transport characteristics, such as those brought about by constructive interference. 

The study of quantum transport in these various open systems has been typically carried out on only a few model systems, and some common network structures; notable exceptions also considering randomly generated networks include Refs.~\cite{Caruso2014UniversallyNetworks, Knee2017Structure-DynamicsSystems, Scholak2011EfficientNetworks, Davidson2021PrinciplesAnalysis}. In the context of biological photosynthetic exciton energy transport this is often the Fenna-Matthew-Olson complex (FMO)~\cite{Olbrich2011TheoryTransitions, Vulto1998ExcitonK, Engel2007EvidenceSystems, Panitchayangkoon2011DirectComplexes, Shabani2012EfficientComplexes}, and in quantum technologies we see (disordered) chains and lattices used to simulate many transport scenarios~\cite{Kassal2012Environment-assistedSystems, Davidson2020TheSubspaces, Coates2021LocalisationTransport, Maier2019Environment-AssistedNetwork, Celardo2016ShieldingHopping, Kropf2019TowardsLayers,Zerah-Harush2020EffectsTransport, Chin2010Noise-assistedComplexes, Mohseni2013GeometricalSystems, Li2015MomentumFlow, Caruso2014UniversallyNetworks}. We see across these contexts that energetic disorder is common in many systems, with the specifics of these energy landscapes having a strong effect on quantum transport~\cite{Davidson2021PrinciplesAnalysis, Knee2017Structure-DynamicsSystems, Coates2021LocalisationTransport,Walschaers2016QuantumUncertainty}. 

Recent work introducing the concept of `population uniformisation' has made the varying transport behaviour between fully coherent and fully classical limits explicit~\cite{Zerah-Harush2018UniversalNetworks}. Population uniformisation states that the variance in on-site populations has a similar qualitative character to the transport efficiency of open quantum systems, and is minimised when transport efficiency is maximised, even in the presence of disorder or repulsive interactions~\cite{Zerah-Harush2018UniversalNetworks,Zerah-Harush2020EffectsTransport}. While this framework explicitly frames things in terms of moving from coherent wavefunctions to classical diffusion and Fick's Law, we yet again see the same, singly peaked ENAQT transport efficiency on the standard systems, including the FMO complex~\cite{Harush2021DoNot}.

In this paper we systematically investigate optimal noise rates across randomly generated transport networks, and show that many have at least two ENAQT peaks or `Goldilocks Zones' where their transport efficiency is maximised. Our networks are made of two-level systems, which we model as point dipoles. We arrange these sites with realistic spacing and effective dipole moments to ensure the relevance of our results, and consider both uniform and normally distributed energy landscapes.

\section{Transport Model}
\label{sec: setup}

\subsection{Network Setup}

\begin{figure}[h]
    \centering
    \includegraphics[width = \linewidth]{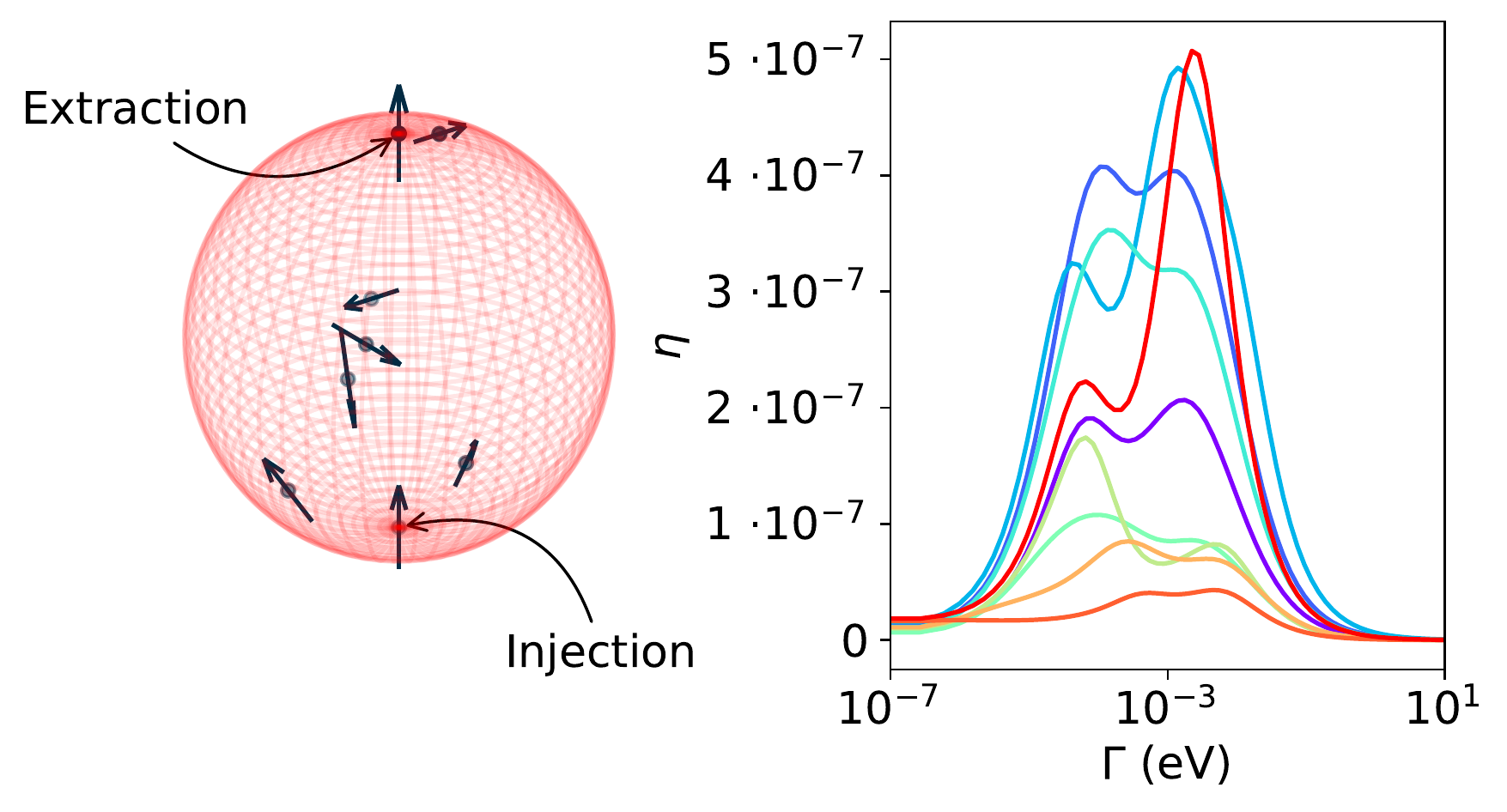}
    \caption{Illustration of a typical dipole system: a fixed volume with injection and extraction sites in at the poles, and randomly oriented and positioned sites within the volume. The system is then coupled to phonon 
environments with varying strengths ($\Gamma$), and the transport efficiency ($\eta$) measured. On the right we show transport efficiency curves from a selection of randomly generated networks modelled with phenomenological pure dephasing that exhibit multiple maxima in their transport efficiency.}
    \label{fig: layout}
\end{figure}

We set up each system as follows: first we define a sphere with a fixed radius of 10 nm, and place two sites at opposite ends of the sphere to act as injection and extraction sites. We then fill the volume with six additional sites with random  positions and dipole orientations to produce a disordered but fully connected network, a similar setup and approach has been utilised in prior works~\cite{Knee2017Structure-DynamicsSystems, Davidson2020TheSubspaces, Walschaers2013OptimallyNetworks}. 
\Cref{fig: layout} shows a typical dipole network generated in this way.

Every site is modelled with an identical dipole moment, we use the effective dipole moments from the bacteriochlorophylls in the FMO complex $|\boldsymbol{d}| = 0.114~ \text{e}\cdot\text{nm} (\sqrt{30}\text{D})$~\cite{Adolphs2006HowBacteria,SchmidtamBusch2011TheProtein,Vulto1998ExcitonK,Wendling2002TheSimulations}. 

With the structure established we can consider the dipole-dipole interactions,
\begin{equation}
    V_{i,j}(\boldsymbol{r}_{i,j}) = \frac{1}{4\pi\epsilon_0} \left( \frac{\boldsymbol{d}_i \cdot \boldsymbol{d}_j}{|\boldsymbol{r}|^3} - 3\frac{(\boldsymbol{r}\cdot \boldsymbol{d}_i)(\boldsymbol{r} \cdot \boldsymbol{d}_j)}{|\boldsymbol{r}|^5}\right) \,,
    \label{eqn: dipole interactions}
\end{equation}
where $\boldsymbol{r}$ is the separation vector between the dipole pair, $\boldsymbol{d}_{i,j}$ are the respective dipole moments and $\epsilon_0$ is the vacuum permittivity. To ensure the point dipole approximation remains appropriate we enforce a minimum separation of 1~nm between every pair of sites~\cite{Krueger1998CalculationMethod}. Throughout this paper we use nanometres, elementary charge and electron volts, meaning a Coulomb constant $\left(4\pi \epsilon_0\right) ^{-1} \approx 1.43996~\text{eV}\cdot\text{nm}~\text{e}^{-2}$. 

To generate the dipole on-site energies we use an average on-site energy $\bar{\epsilon} = 1.5498$~eV (12,500~cm$^{-1}$), and will later also sample them from a normal distribution with a standard deviation of 1\%, so $\sigma$ = 0.0155eV, very similar to the energies and disorder found in the FMO complex~\cite{Adolphs2006HowBacteria}. The default values we use are summarised in \Cref{tab: dipole network properties}.

\begin{table}[H]
\centering
\caption{Properties used to generate the main network ensemble. The rationale behind our parameter choices is given in the main text.}
\begin{tabular}{c c c c c c}
\toprule
     Radius(nm) & N & $r_{min}$ (nm)& $|\boldsymbol{d}|$ (e $\cdot$ nm) & $\bar{\epsilon}$ (eV) & $\sigma$ (eV) \\
     \midrule
     10 & 8 & 1 & 0.11403 &  1.5498 & 0.0155 \\
\bottomrule
\end{tabular}

\label{tab: dipole network properties}
\end{table}

With the on-site energies and dipole-dipole interactions defined we can construct the Hamiltonian. We assume there is only a single excitation in the system at any time and construct the following excitonic Hamiltonian 
\begin{equation}
    H =\sum^N_i \epsilon_i \ket{i}\bra{i} + \sum_{i \neq j}^N V_{i,j} \ket{i}\bra{j} \,,
    \label{eqn: Hamiltonian}
\end{equation}
where $i$ and $j$ are site indices, and the $V_{i,j}$ are evaluated according to \cref{eqn: dipole interactions}. We use the single excitation approximation for computational efficiency as it gives an $ N \times N$ Hamiltonian, and is widely used when modelling open quantum systems with low injection rates, such as light harvesting complexes~\cite{Blankenship2002MolecularPhotosynthesis, Davidson2020TheSubspaces, Walschaers2013OptimallyNetworks, Knee2017Structure-DynamicsSystems, Ishizaki2009UnifiedApproach, Moix2011EfficientFMO, Shabani2012EfficientComplexes}.

\subsection{Dynamics}
\label{sec: theory}

To look at the effect of the environment and finite temperatures on transport properties, we use the full non-secular Bloch-Redfield master equation~\cite{Breuer2007TheSystems, Jeske2015Bloch-RedfieldComplexes}
\begin{equation}
    \begin{split}
        \Dot{\rho_s} =& -i [H,\rho_s] \\
        &+ \gamma_{inj} \mathcal{D}\left[A_{inj} \right]\rho_s
        + \gamma_{ext}\mathcal{D}\left[A_{ext}\right]\rho_s\\ 
        &+ \sum_{\omega} \sum_{m, n} S_{m,n}(\omega) ( A_n(\omega)\rho_s A^\dagger_m(\omega) \\
        &- \frac{1}{2} \{ A^\dagger_m(\omega)A_n(\omega), \rho_s \} ) \,,
    \label{eqn: redfield ME}
    \end{split}
\end{equation}
where $\rho_s$ is the system Hamiltonian and $\omega$ are the eigenenergy splittings~\cite{Breuer2007TheSystems, Jeske2015Bloch-RedfieldComplexes}. The system-environment interactions, $A_{m}$, are derived by transforming the relevant site basis operators $A_{deph,i} = 2\ket{i}\bra{i} - \mathbb{I}$ into the Hamiltonian eigenbasis and $S_{m n}(\omega)$ defines the noise-power spectrum associated with the system-environment interaction~\cite{Breuer2007TheSystems, Jeske2015Bloch-RedfieldComplexes, Davidson2020TheSubspaces}. The noise-power spectrum function is

\begin{equation}
    S_{m, n}(\omega) =\left(\mathcal{N}_{BE}(\omega, \beta) + \Theta(\omega)\right)\mathcal{J}(\omega),
    \label{eqn: noise-power-spectrum}
\end{equation}

where $\mathcal{N}_{BE}(\omega)$ defines Bose-Einstein statistics at a given phonon inverse temperature $\beta$, $\Theta(\omega)$ is the Heaviside function, allowing phonon-assisted transitions from higher to lower eigenenergies ($\omega > 0$), and $\mathcal{J}(\omega)$ is the phonon spectral density~\cite{Jeske2015Bloch-RedfieldComplexes}. In this work we use the Drude-Lorentz spectral density, which has previously been used to model excitonic transfer in light harvesting complexes~\cite{Fassioli2012CoherentConditions, Kreisbeck2014ScalableModes}, 
\begin{equation}
    \mathcal{J}(\omega) = \Gamma \cdot \frac{2}{\pi} \cdot \frac{\omega (1/\tau)}{\omega^2 + (1/\tau)^2},
    \label{eqn: drude-lorentz-functions}
\end{equation}
where $\Gamma$ scales the rate of noise in the system from interactions with the environment, $\tau$ is the correlation time, and $\tau^{-1}$ is the spectral density peak frequency.

Finally, we have \(\mathcal{D}\left[A\right]\rho \) which is the dissipator superoperator

\begin{equation}
    \mathcal{D}\left[A\right]\rho = \left(A \rho A^\dagger - \frac{1}{2} \{ A^\dagger A, \rho \} \right) \,.
    \label{eqn: dephasing superop}
\end{equation}

To model extraction and injection, a shelf state is appended to the system. The extraction operator $A_{ext}$ projects population from the extraction site to the shelf state, 
$A_{ext} = {\sigma_{shelf}^+\sigma_{ext}^-}$, and then that population is re-injected from the shelf state back onto the injection site with the injection operator $A_{inj} = {\sigma_{inj}^+\sigma_{shelf}^-}$. Injection and extraction are matched, $\gamma_{ext, inj} = 0.1~\text{eV}$, changing this value generally changes quantitative values but not the qualitative behaviour ~\cite{Zerah-Harush2018UniversalNetworks}.

To complement the Redfield calculations, we also carry out phenomenological pure dephasing calculations with the Lindblad master equation~\cite{Breuer2007TheSystems, Lindblad1976OnSemigroups,Gorini2008CompletelySystems}

\begin{equation}
    \begin{split}
        \Dot{\rho} =& -i [H,\rho] 
        + \Gamma \sum_{i = 1}^N \mathcal{D}\left[A_{deph, i}\right]\rho \\
        &+ \gamma_{inj} \mathcal{D}\left[A_{inj} \right]\rho
        + \gamma_{ext}\mathcal{D}\left[A_{ext}\right]\rho \,, 
    \end{split}
    \label{eqn: lindblad ME}
\end{equation}
where $A_{deph,i}$ are Lindblad operators describing the environmental influence on each site $i$, again these are of the form $A_{deph,i}=2\ket{i}\bra{i}-\mathbb{I}$~\cite{Jeske2015Bloch-RedfieldComplexes}. All other symbols have the same meaning as in \cref{eqn: redfield ME}. This approach is equivalent to the nonsecular Bloch-Redfield master equation for an infinite temperature and a flat spectral density~\cite{Davidson2021PrinciplesAnalysis}.

We focus here on the steady state $\rho_{ss}$ which is found by calculating the null vector of the system evolution Liouvillian. Our figure of merit then is the excited steady state population on the extraction site
\begin{equation}
    \eta = \bra{\text{extraction}}\rho_{ss}\ket{\text{extraction}}.
    \label{eqn: extraction population}
\end{equation}

This is motivated by the strong correspondence found in prior work between dynamical and steady state transport properties~\cite{Zerah-Harush2020EffectsTransport, Dubi2015InterplayComplexes}, as well as further studies suggesting that the steady state is more natural for photosynthetic systems~\cite{Brumer2018SheddingProcesses, Axelrod2018AnLight, Kassal2013DoesPhotosynthesis}. The steady state approach also allows us to avoid any confusion that could arise from the influence of transient effects when comparing different networks.

For each network considered in this paper, we were interested in how this transport efficiency $\eta$ changes with $\Gamma$, the noise rate from coupling to the environment. To do this we considered a large range of noise rates $\Gamma~\mapsto~\left[10^{-7} - 10~\text{eV}\right]$, and for each value recorded $\eta$ as well as the full steady state population. This range of $\Gamma$ was chosen as it was broad enough to capture the values where $\eta$ has maxima for our networks, and additionally show the transport efficiency decreasing outside these peaks as shown in \cref{fig: layout}. 

These results were then filtered to ensure validity, all data presented here has passed checks on the unity of the steady state trace, non-negativity of on-site populations and steady state eigenvalues (see \cref{sec: filtering}). From that point we could perform simple peak--finding calculations for each network and spectral density to directly identify in which cases there was more than one optimal noise rate or peak in the transport efficiency curves, and how often this occurred. 

\section{Results}
\label{sec: results}

\subsection{Environmental Effects}
We generated several ensembles of 1,000 dipole networks, each comprising two fixed and six randomly located sites. For our Redfield calculations the key spectral density parameters are the temperature, and the peak frequency  of the Drude-Lorentz spectral density ($\tau^{-1}$). Lindbladian pure dephasing acts as our infinite temperature limit, being equally present at all frequencies. 

We start by considering an ensemble of networks with identical splitting between the two levels on each site, and see how many networks have multiple maxima in their transport efficiency. This approach lets us compare our results to prior works that have made the same assumption of uniform on-site energies when modelling disordered molecular networks and other complexes with dipole interactions~\cite{Mostarda2013StructuredynamicsSystems, Walschaers2013OptimallyNetworks, Scholak2011EfficientNetworks}.

\begin{figure}[H]
    \centering
    \includegraphics[width = \linewidth]{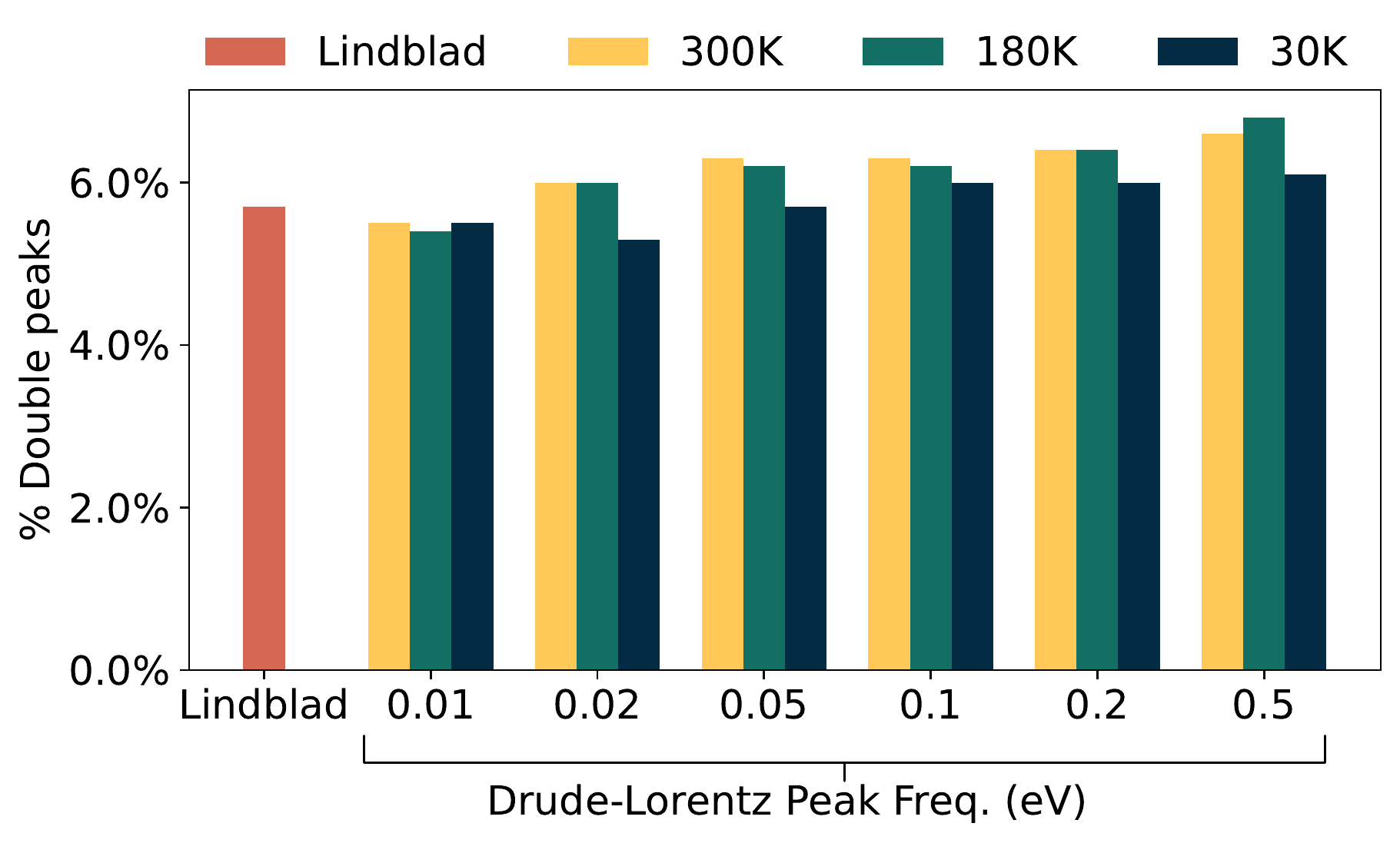}
    \caption{Percentage of dipole networks showing double-ENAQT behaviour in networks with uniform on-site energies. We see only a slight sensitivity to the peak frequency in this case, and a slight increase in the behaviour at low temperatures. This is likely because the eigenenergy splittings are quite small when the on-site energies are identical, and as such the transport behaviour depends on the near zero behaviour of the Drude-Lorentz spectral density, which is a relatively stable region.}
    \label{fig: double enaqt degenerate bar chart}
\end{figure}

\begin{figure}[H]
    \centering
    \includegraphics[width = \linewidth]{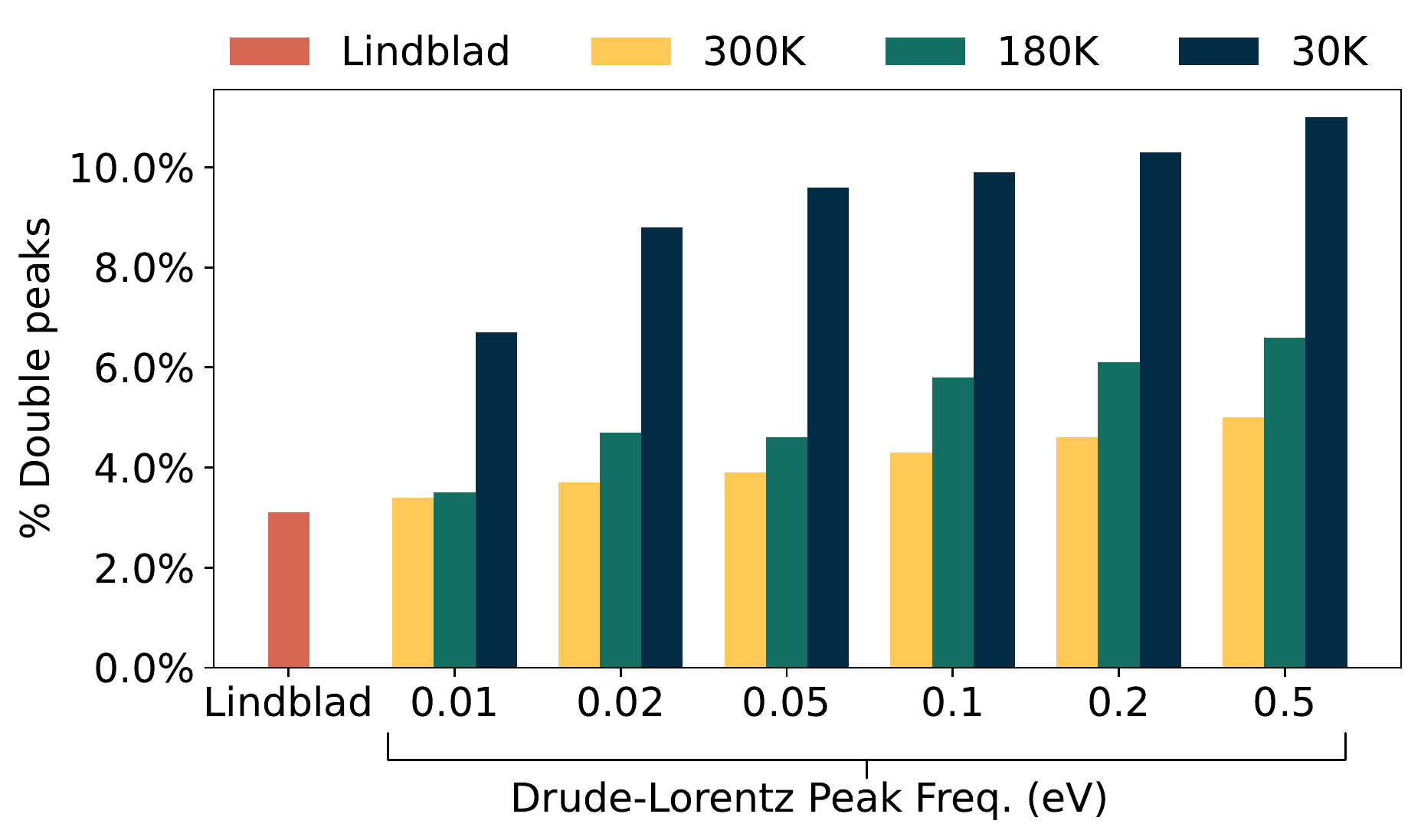}
    \caption{Percentage of networks with normally distributed energies showing double-peaked ENAQT behaviour. The lower the temperature and the higher the environment peak frequency, the more often we observe this phenomenon. There is a much larger sensitivity to the spectral density than in the energetically uniform case.}
    \label{fig: double ENAQT bar chart}
\end{figure}

\Cref{fig: double enaqt degenerate bar chart} shows a surprising result, contrary to prior wisdom we consistently find about 6\% of networks have multiple peaks in their transport efficiency, regardless of the spectral density or temperature. This illustrates that these fully connected networks can have multiple maxima in their transport efficiency, but neglects the importance of on-site energies in transport~\cite{Davidson2021PrinciplesAnalysis}. We consider the effects of varied on-site energies in \cref{fig: double ENAQT bar chart}, where the energies are normally distributed as described in \cref{tab: dipole network properties}. 

\Cref{fig: double enaqt degenerate bar chart} and \cref{fig: double ENAQT bar chart} show our main results from this paper. We see that in every situation we simulate, a sizeable subset of our networks have two peaks in the their transport efficiency. The Drude-Lorentz peak frequency has a slight effect on how often we observe this behaviour, but there is a more pronounced sensitivity to temperature. The lower the temperature, the more often we see this behaviour. By extension, this is seen least often -- but still clearly represented -- in the Lindblad pure dephasing limit. We show what proportion of these results occur within measured FMO reorganisation energies in \cref{sec: reorganisation energy}.

We note that this double peaked phenomenon generally occurs more frequently in the energetically uniform ensemble. We attribute this to energetic disorder producing greater localisation. Meaning that not only are the energetic differences between eigenstates larger, but also those eigenstates are more tightly confined to specific sites. This greater spatial confinement means there are fewer pathways from injection site to extraction site. The greater energetic differences also raise the chance of some eigenstates being so far detuned that they are effectively inaccessible given the finite range of noise rates $\Gamma$ we consider.

\subsection{Structural Effects}
\label{sec: structural effects}

As shown on the right of \cref{fig: layout}, different networks have transport efficiency curves that take a variety of forms. The relative prominence and position of each transport efficiency peak varies. In some cases peaks are very well separated with a pronounced dip between them. In other cases the two are barely distinguishable, almost smearing into each other. This suggests a strong structural dependence between the form of the curves and the Hamiltonians of their respective networks. 

\begin{figure}[H]
    \centering
    \includegraphics[width = \linewidth]{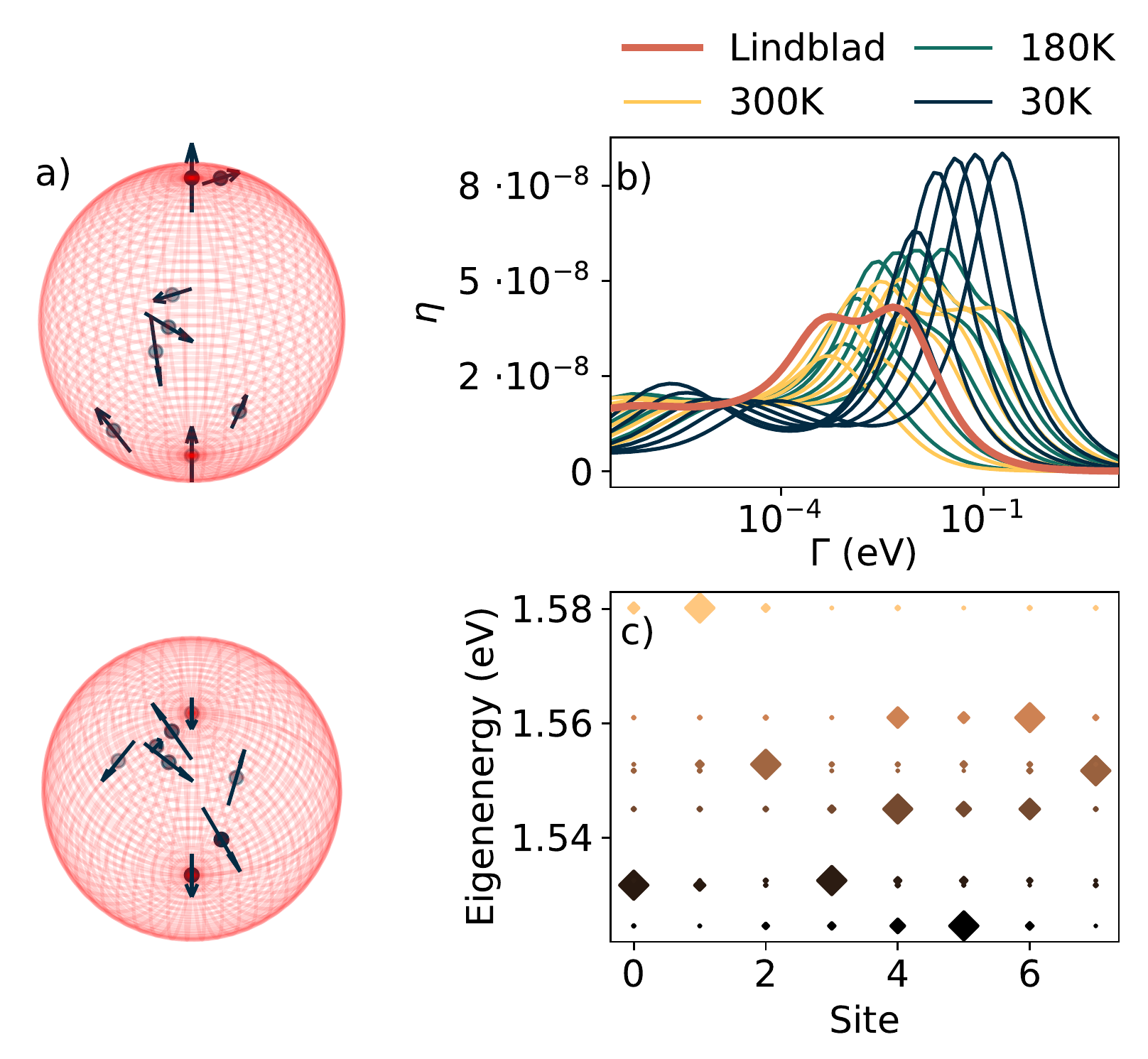}
    \caption{A system where we see double-peaked ENAQT for all forms of spectral density we consider. (a) shows the dipole positions and orientations for this network, with the $z$-axis pointing upwards, and then out of the page (tilted by $\pi/2$) in the top and bottom plot, respectively. (b) shows the transport efficiency in all our cases and (c) shows the network Hamiltonian eigenstates, with eigenenergies on the $y$-axis and site indices on the $x$-axis (0 and 1 are the injection and extraction sites, respectively). The size of the diamonds indicate the support of the eigenstate on that site.}
    \label{fig: multipeak network}
\end{figure}

We closely inspected networks such as the one in \Cref{fig: multipeak network}, which produce double peaks across a wide range of temperatures and Drude-Lorentz peak frequencies, to identify what key features might be correlated with having multiple peaks in the transport efficiency of a system. See \cref{sec: specific hamiltonians} for the physical properties of the network.

We did not find any strong geometric dependence across networks with multiple maxima in their transport efficiency, but we identified consistent features in the system eigenstates. Specifically, these systems often have one or more large gaps in their eigenenergy distributions, as shown in \cref{fig: multipeak network} (c). 

\subsection{Discussion}
\label{sec: summary}

An intuitive and tempting interpretation of the multiple optimal transport regimes would be that two broad sets of energy scales are important: the width of each band of eigenenergies, and the separation between bands. The intuition from this being that there is an initial path with an ideal environmental coupling for ENAQT. However, as phonon couplings become larger, eigenstates that were previously far--detuned become more accessible. If the new path is more efficient for exciton transport, then a new peak in the transport efficiency can be observed. Both the energetically uniform and non-uniform ensembles consistently have these gaps in their spectra, induced by the dipole-dipole interactions between sites, and if present, differences in on-site energies.

Testing this hypothesis in \cref{sec: ensemble eigenenergies} we do indeed find a positive correlation with the relative standard deviation of eigenenergy splittings. However the change in the fraction of multiply peaked networks remains modest, suggesting there are other factors at play. In the following we briefly summarise how doubly peaked behaviour correlates with some other aspects.

In \cref{sec: offset networks} we tested the energetic separation hypothesis in another way, generating a new independent ensemble of 1000 networks where three of the dipoles have a fixed offset added to their on-site energies. This approach encourages more gaps to form in the eigenspectra and we see a modest increase in the frequency of double peaks.

We also have considered the relative energies of the injection and extraction sites in these systems. In \cref{sec: energy direction} we show that double peaks occur more frequently when injecting at lower energies than the extraction site, suggesting it may occur more often in less efficient networks (in the sense of requiring `uphill' energy transport), albeit by no means limited to those. In \cref{sec: single vs double peak efficiency} we directly compare the maximum transport efficiency of single-peaked and double-peaked networks, and show that double-peaked networks have a larger spread in their  transport efficiencies, but can be just as efficient as the single-peaked networks. \Cref{sec: relative peak efficiency} is then concerned with how relevant each peak is in double-peaked systems. We find that that both peaks typically have a similar prominence, though the peak at higher system-environment couplings tends to be more efficient. 

Another consideration is the number of potential paths in a system from the injection site to the extraction site. We generated an ensemble of networks made of paired, disordered, nearest-neighbour chains that only connected at shared injection and extraction sites, giving only two paths across the system. In \cref{sec: double chains} we show that while this strongly reduces how often a network has multiple transport efficiency maxima, we do still observe it against all spectral densities. We further show in \cref{sec: ohmic superohmic} that double peaks can be observed against Ohmic and superohmic spectral densities as well. To consider the effect of system density, in \cref{sec: dense-network} we reduce the minimum separation between sites and the total system volume to better match the chromophoric density seen in light-harvesting complexes. Again we find a similar subset of networks with multiple optimal noise rates, though one that less favours multiple ENAQT peaks at low temperatures.

Overall, our analysis suggests there are a multitude of factors at play which can positively correlate with an increased occurrence of doubly peaked networks. The analysis in this paper has been focused on networks with double peaks as that is what we observe for these systems. We believe that more than two ENAQT peaks are possible, and that networks with more sites and potential paths from source to sink may present such behaviour. Given the large amount of parameters involved in these systems we have presented many conditions that allow for these multiple peaks to occur in a range of system geometries, but do not find any condition that strongly correlates with the multiple ENAQT peaks being present.

\section{Conclusion}
\label{sec: conclusion}
In this paper, we have shown that transport networks with realistic and microscopically resolved vibrational interactions frequently feature more than one optimal regime for transport efficiency. This runs counter to expectations of there being a single `Goldilocks zone' in any and all cases. We observe that these multiple optimal transport regimes can occur for energetically ordered or energetically disordered networks, and occur more often in networks with less evenly spaced eigenvalues.

\section*{Conflicts of interest}
There are no conflicts to declare.

\section*{Acknowledgements}
This work was supported by EPSRC Grant Nos. \\EP/L015110/1, EP/T007214/1, EP/T01377X/1, and EP/T014032. Calculations carried out in QuTiP~\cite{Johansson2013QuTiPSystems}, data analysis carried out with pandas DataFrames~\cite{Mckinney2010DataPython, Thepandasdevelopmentteam2021Pandas-dev/pandas:1.3.4}.








\bibliography{references}  
\bibliographystyle{rsc}

\appendix
\section{Role of eigenenergy spacing disorder}
\label{sec: ensemble eigenenergies}

\begin{figure}[H]
    \centering
    \includegraphics[width = \linewidth]{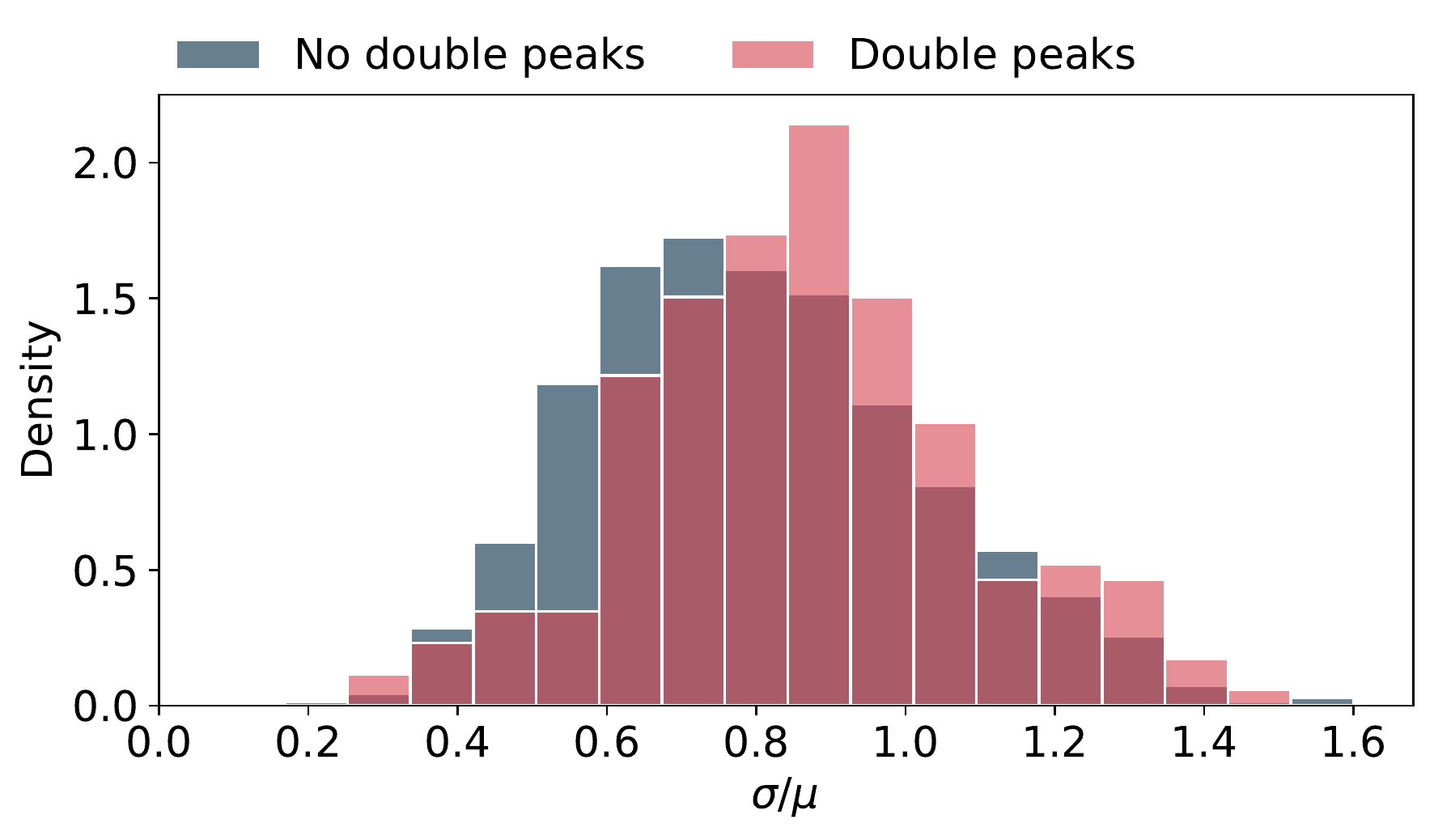}
    \caption{Histogram of probability density against the relative standard deviation of network eigenenergy differences for energetically disordered networks. We see the networks with double peaks occur more often for higher amounts of disorder. We test this directly in \cref{sec: offset networks}.}
    \label{fig: hist-std-normal}
\end{figure}

\begin{figure}[H]
    \centering
    \includegraphics[width = \linewidth]{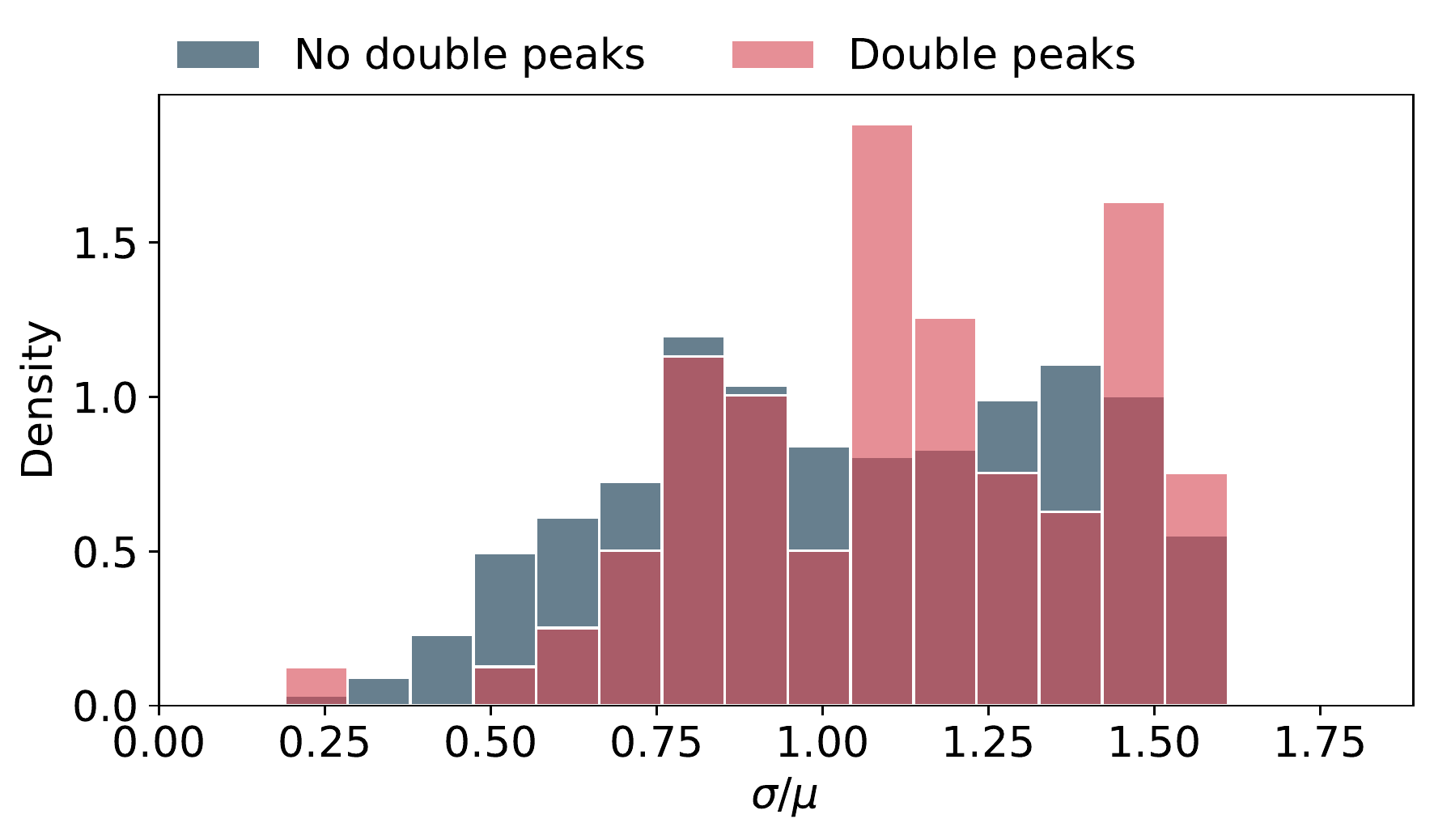}
    \caption{Histogram of probability density against the relative standard deviation of network eigenenergy differences for 1,000 energetically uniform networks, previously described in \cref{fig: double enaqt degenerate bar chart}. Without an energy landscape, we see a wide spread of relative standard deviations defined by the geometric properties of the networks. We see a relatively sharp maximum relative standard deviation here due to the exclusion volume or minimum distance we enforce between dipoles when generating our systems.}
    \label{fig: hist-std-degenerate}
\end{figure}

Here we present statistical analysis of the role of disorder in the spacing of eigenenergies, specifically we consider the relative standard deviation $\frac{\sigma}{\mu}$ where $\sigma$ is the standard deviation of the eigenenergy differences, and $\mu$ is the average eigenenergy difference. We find in both our energetically uniform and energetically disordered ensembles that more disorder in eigenenergy spacings is positively correlated with an increased fraction of networks displaying two optimal transport regimes. The probability density histograms against the relative standard deviation are shown for the energetically disordered and energetically uniform ensembles in \cref{fig: hist-std-normal} and \cref{fig: hist-std-degenerate} respectively.

\section{Networks with offset energies}
\label{sec: offset networks}
Based on our intuition of double peaks being related to two effective pathways with different optimal conditions for ENAQT, we generate a new, independent ensemble of networks with two different on-site energy scales. We generate the dipole networks exactly as before, including their random on-site energies. Then for each network we randomly pick three of the bulk dipoles and shift their energy upwards by the standard deviation of our on-site energies (+15.5~meV). 

\begin{figure}[H]
    \centering
    \includegraphics[width = \linewidth]{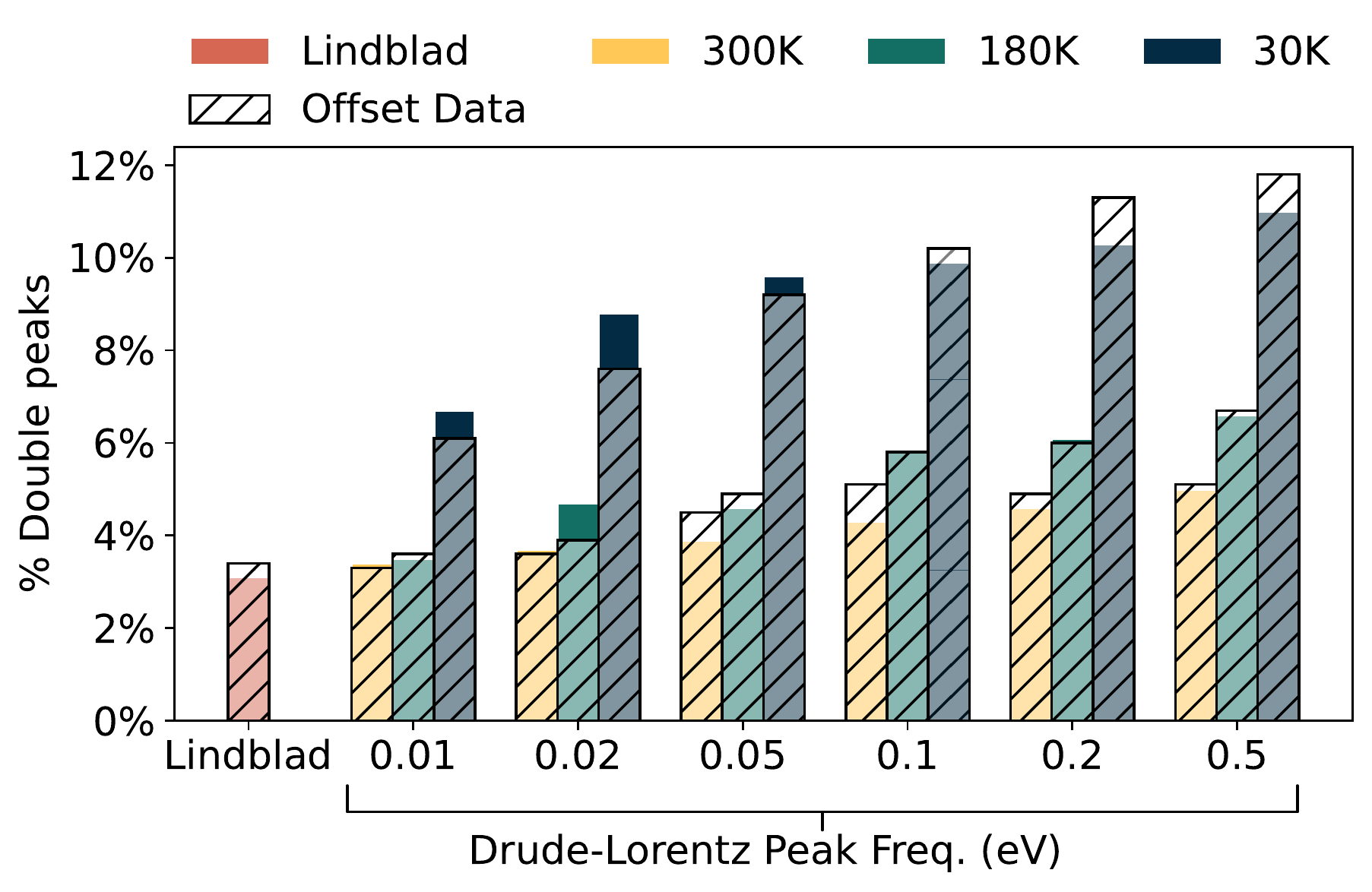}
    \caption{The hatched bars show percentage of dipole networks with double-ENAQT behaviour in our artificially offset networks. The coloured bars show the prior results from \cref{fig: double ENAQT bar chart} for the normally distributed on-site energies. We see a general increase in double peaked behaviour thanks to this energy offset. }
    \label{fig: artificial network}
\end{figure}

This adjustment to the networks increases the probability of there being a larger gap in the system eigenenergies, but leaves the geometric properties of these networks unaffected. As shown in \cref{fig: artificial network}, we do see some increase in double-peaked transport efficiency in most cases. Though as expected when comparing two independent datasets there are fluctuations in the trends. 

\section{Injection and extraction energies}
\label{sec: energy direction}

We also consider the importance of where the injection and extraction sites sit in the eigenenergy landscape. We index the eigenstates from lowest to highest energy ($\lambda_{1 \longrightarrow N}$), and record which eigenstate is most present on the injection site ($\lambda_{inj}$) and the extraction site ($\lambda_{ext}$), respectively. 

\begin{figure}[H]
    \centering
    \includegraphics[width = \linewidth]{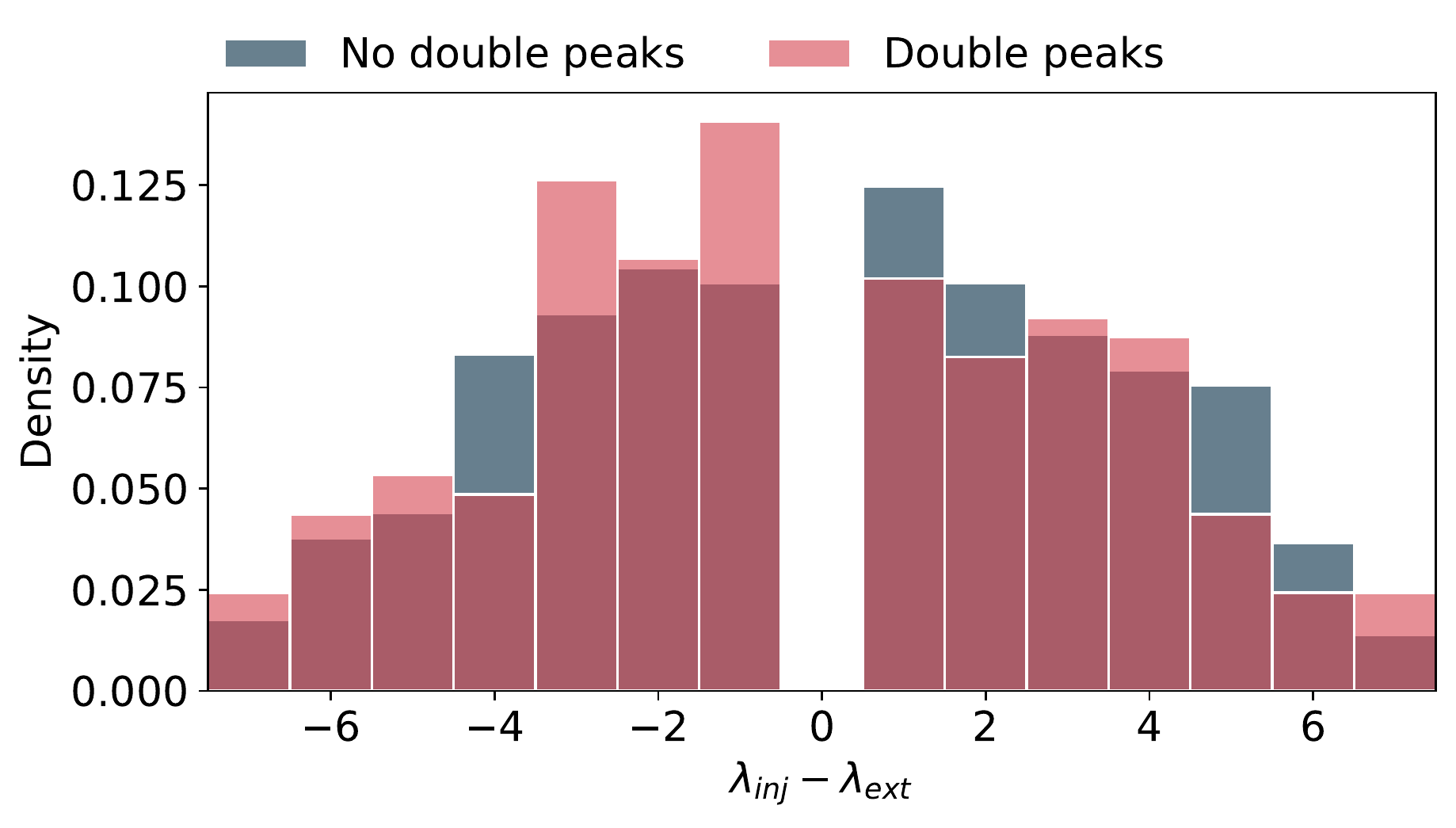}
    \caption{Histogram of probability density against the difference in eigenenergy index of the injection and extraction sites for energetically disordered systems.  Negative values mean the injection is below the extraction. }
    \label{fig: hist-kde-inj-ext-normal}
\end{figure}

\begin{figure}[H]
    \centering
    \includegraphics[width = \linewidth]{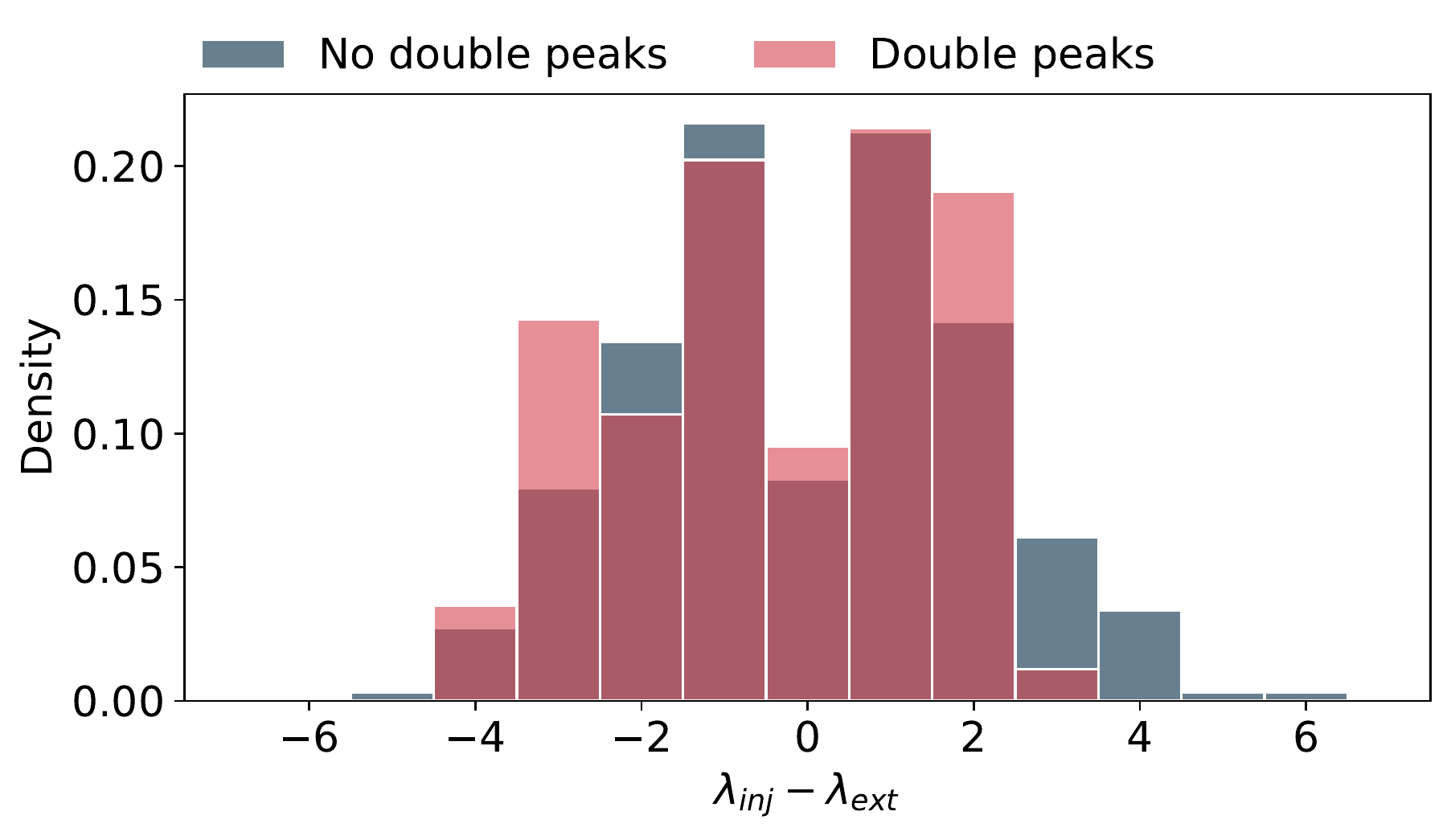}
    \caption{Histogram of probability density against the difference in eigenenergy index of the injection and extraction sites. Negative values mean the injection is below the extraction. We again see fewer cases where doubly peaked networks have the injection far above the extraction, though the general trend is less clear than for the energetically disordered case (\cref{fig: hist-kde-inj-ext-normal}). }
    \label{fig: hist-kde-inj-ext-degenerate}
\end{figure}

For our networks with no double peaks, there is general symmetry. For the networks with multiple transport efficiency maxima, we see a preference for injecting at lower eigenenergies than they extract at ($\lambda_{inj} - \lambda_{ext} < 0$). This suggests the effect will be more prominent in less efficient systems. Though we note that the effect is present broadly, also clearly occurring in networks where energy transport should be efficient along a downhill gradient. \Cref{fig: hist-kde-inj-ext-normal} shows these results for energetically disordered networks, and \cref{fig: hist-kde-inj-ext-degenerate} shows the same for energetically uniform networks. We see the same trend in both results, however for the energetically uniform case the eigenstates are more delocalised, often being spread over two or more sites. As a result there are multiple cases where $\lambda_{inj} - \lambda_{ext} = 0$ due to the sites sharing a pair of eigenstates which are equally present on the injection and extraction sites.  This does not occur in the energetically disordered ensemble because of the additional localisation.

\section{Double peaks in coupled chains}
\label{sec: double chains}

To further examine the importance of different energy scales, we constructed a simple system from 8 sites. We define two nearest neighbour chains of three sites, and connect them to an injection site and extraction site at either end. Both chains have the same NN coupling (2.5~meV) and average on-site energy (1.55~eV), but have different amounts of disorder in on-site energies. We set their standard deviations to $\sigma = 15.5 \text{ and } 1.55$~meV, respectively, such that each arm will typically have a different amount of phonon coupling that is optimal for overcoming localisation. 

\begin{figure}[H]
    \centering
    \includegraphics[width = \linewidth]{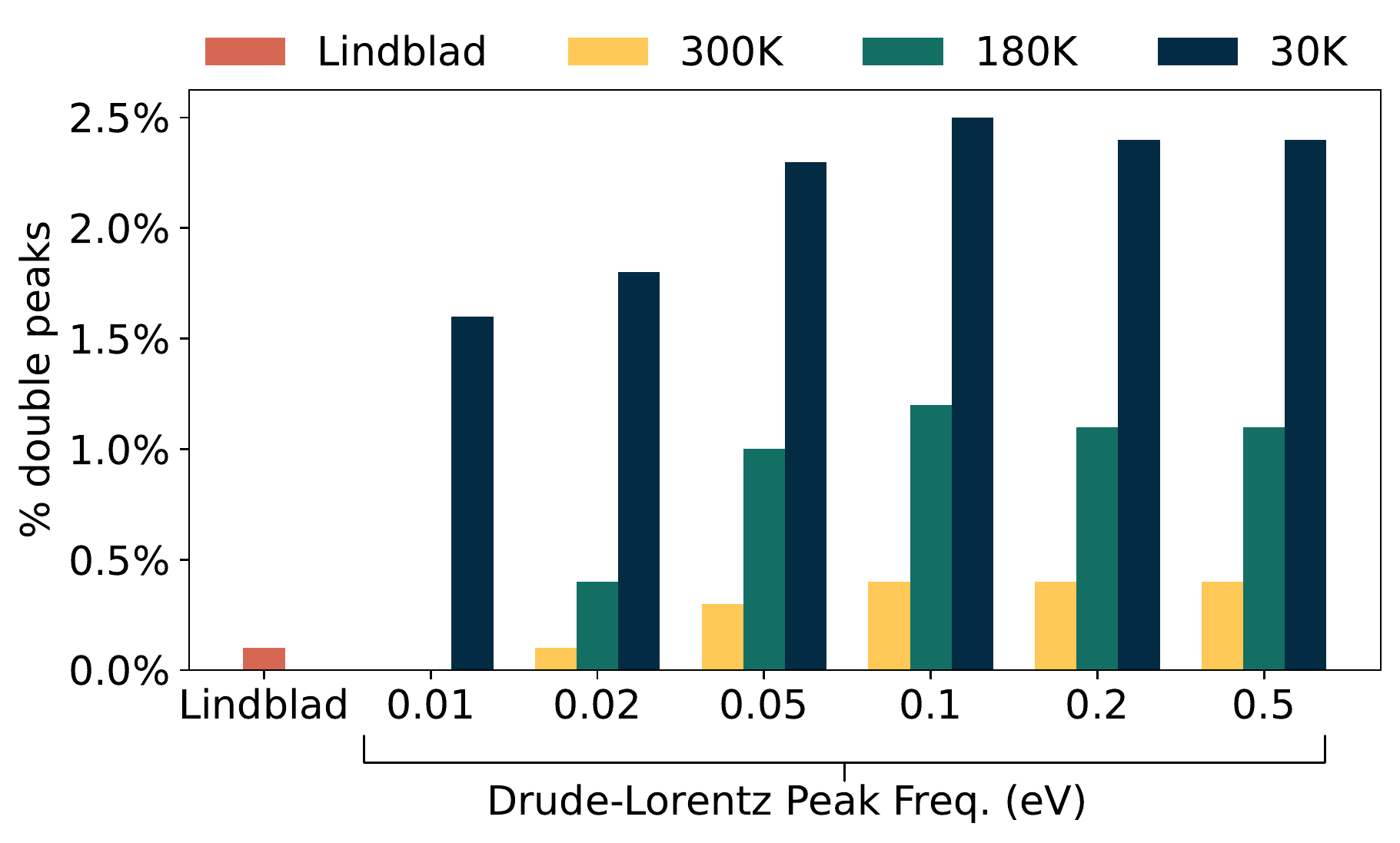}
    \caption{Double-peak rates for 1000 `two-armed' networks. We see a decrease in double-peak behaviour in every circumstance compared to our main results shown in \cref{fig: double ENAQT bar chart}. This suggests that the reduction of available paths or long-range coupling is limiting how often double peaks occur.}
    \label{fig: two-arm enaqt}
\end{figure}

This gives us a scenario with two well-separated paths between injection and extraction, rather than the many possible traverses in our dipole networks. As such, there is a general reduction in secondary pathways that have an opportunity to improve transport efficiency. The results of these calculations for 1,000 networks are shown in \cref{fig: two-arm enaqt}.

As \cref{fig: two-arm enaqt} shows, double peaks are still present, though always to a lesser degree than in our totally random dipole ensembles. A key difference is the large decrease of double peaks with Lindblad pure dephasing, or at high temperatures but low peak frequencies. So just having two possible pathways across a system is not enough to remove the possibility of double peaks, but does lower the chances of observing it.

\section{Other spectral densities}
\label{sec: ohmic superohmic}
The arguments presented in this paper are not unique to the Drude-Lorentz spectral density or pure dephasing. We can also consider the Ohmic and superohmic spectral densities. We define these spectral densities with Gaussian cutoffs as
\begin{equation}
    \mathcal{J}(\omega) = \Gamma \cdot \left(\frac{\omega}{\omega_c}\right)^S e^{-\left(\frac{\omega}{\omega_c}\right)^2},
    \label{eqn: ohmic-functions}
\end{equation} 
where $\Gamma$ is the system-phonon coupling, and $\omega_c$ is the cutoff frequency. $S = 1 \text{ and } 3$ for our Ohmic and superohmic powers respectively. The peak frequency $\omega_{peak}$ depends on the cutoff as 
\begin{equation}
    \omega_{peak} = \omega_{c} \sqrt{\frac{S}{2}}.
\end{equation}

\begin{figure}[H]
    \centering
    \includegraphics[width = \linewidth]{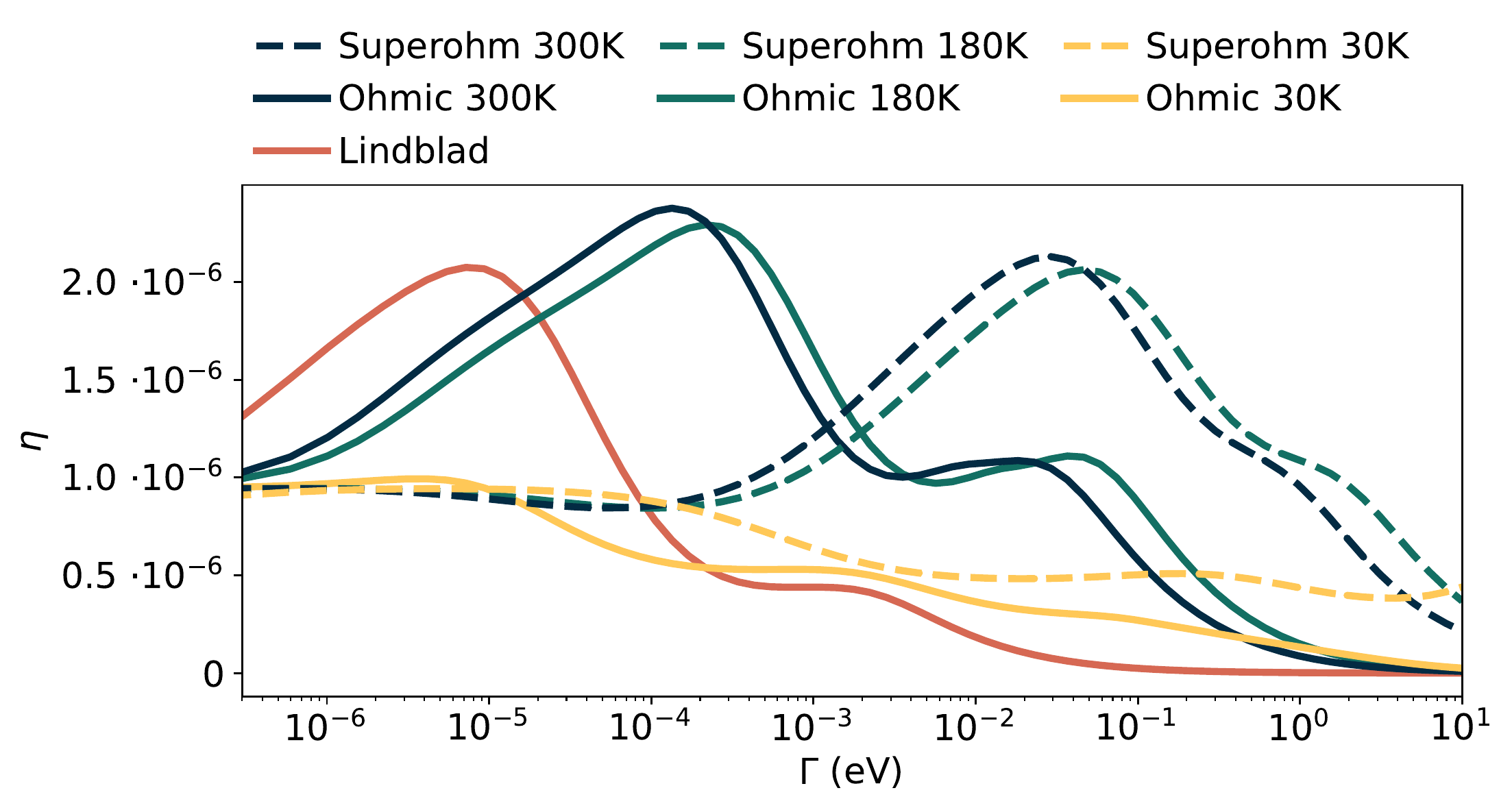}
    \caption{Multi-peaked transport efficiency curves for a single network with uniform on-site energies, showing results from Lindblad pure dephasing, as well as Ohmic and superohmic spectral densities with  $\omega_{peak} = $ 0.1~eV. We see similar qualitative behaviour across the Lindblad, 300~K and 180~K curves, with a change in behaviour at 30~K.}
    \label{fig: power law double peak}
\end{figure}

We observe that when secondary peaks appear in these scenarios, they are often at very high values of $\Gamma$, compared to the range we use to see the same behaviour in pure dephasing and Drude-Lorentz models. This is slightly mitigated in the energetically uniform networks where the lack of disorder in  on-site energies has the effect of moving these peaks to lower coupling strengths where our standard approach can capture them. We present a clear example in \cref{fig: power law double peak} of a single energetically uniform network showing double peaks at all temperatures tested for the Ohmic and superohmic distributions.

As such we can state that these double-peaked effects can and do also occur for these power law spectral densities. However, they occur over a much broader range of environmental couplings, and as such, alternative methods suited to intermediate- and strongly-coupled open quantum systems would be needed to provide more robust statistics.

\section{Hamiltonian Details}
\label{sec: specific hamiltonians}

Here we list out the data for the specific dipole network presented in \cref{fig: multipeak network}. 
\begin{table}[H]
\centering
\caption{Dipole positions and Energies}
\begin{tabular}{lllll}
\toprule
Dipole &    X (nm) &    Y (nm) &    Z (nm) & Energies (eV)\\
\midrule
0 (inject) & 0.0 &       0.0 &     -10.0 &      1.552794 \\
1 (extract) &  0.0 &0.0 &  10.0 &      1.524548 \\
2 & -5.239018 & -2.013063 & -6.763873 & 1.544986 \\
3 & -2.429034 &  1.463867 & -2.933762 & 1.552236 \\
4 & -1.321062 &  4.226071 & -0.255611 & 1.580151 \\
5 &  3.148822 & -2.374797 & -5.102531 & 1.532472 \\
6 & -1.552033 &  1.351976 & -1.062326 & 1.560427 \\
7 &  1.851469 &  1.995554 &   9.06525 & 1.53166 \\
\bottomrule
\end{tabular}
\label{tab: dipole positions}
\end{table}

\begin{table}[H]
\centering
\caption{Dipole vector x, y and z components.}
\begin{tabular}{llll}
\toprule
Dipole & $\boldsymbol{d}_x$ (e $\cdot$ nm) & $\boldsymbol{d}_y$ (e $\cdot$ nm) & $\boldsymbol{d}_z$ (e $\cdot$ nm) \\
\midrule
0 &       0.0 &       0.0 &  0.114033 \\
1 &       0.0 &       0.0 &  0.114033 \\
2 & -0.054907 & -0.023364 &  0.097174 \\
3 &  0.016014 & -0.040995 & -0.105197 \\
4 & -0.065844 &  0.071077 & -0.060133 \\
5 &   0.02771 &  0.110455 & -0.005934 \\
6 &  0.081605 & -0.079198 & -0.008472 \\
7 &  0.058304 & -0.079757 &  0.056948 \\
\bottomrule
\end{tabular}
\label{tab: dipole moments}
\end{table}

\section{Reorganisation Energies}
\label{sec: reorganisation energy}

\begin{figure}[H]
    \centering
    \includegraphics[width = \linewidth]{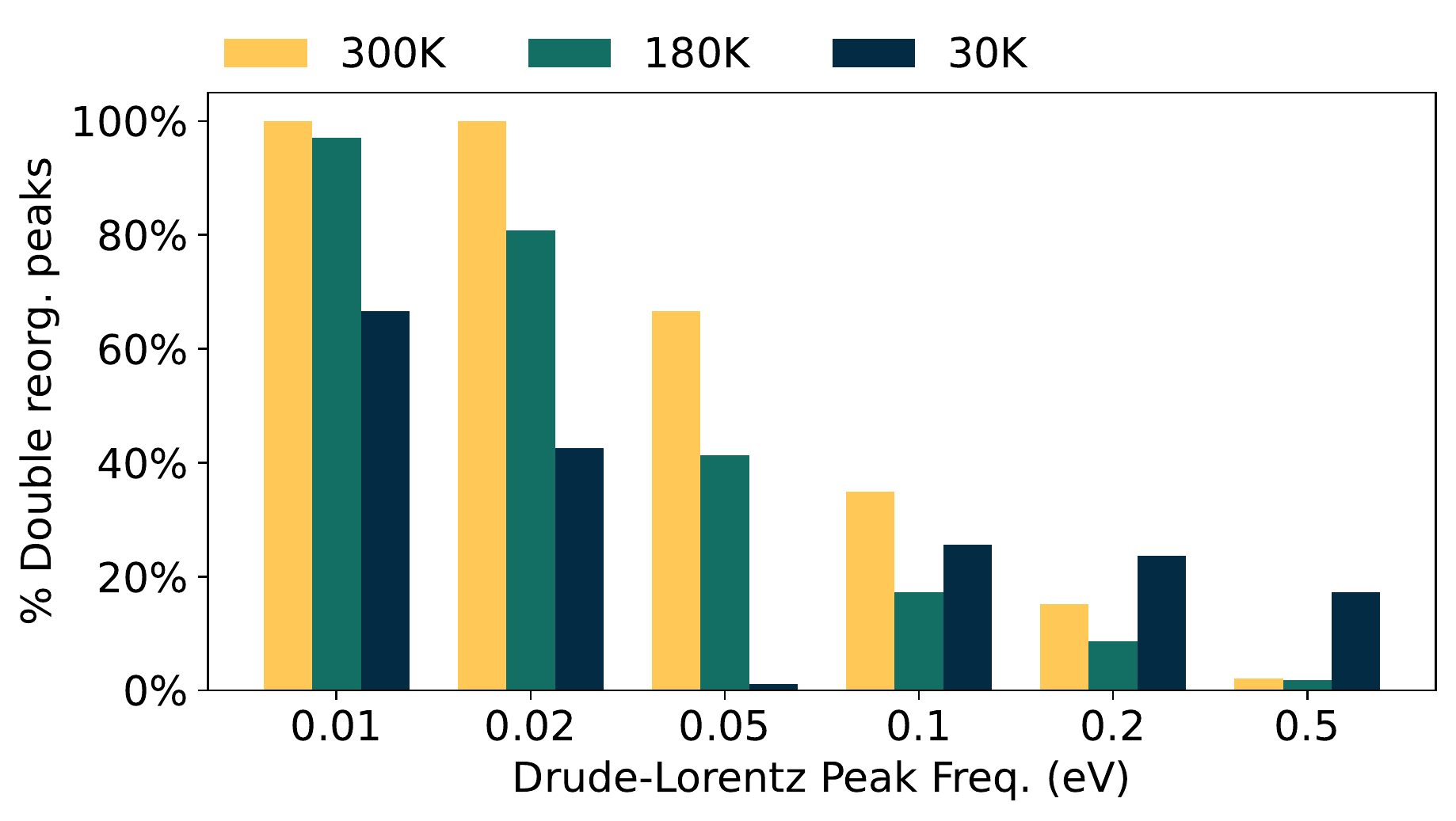}
    \caption{Bar chart showing what percentage of double-peaked networks have both their peaks below a maximum reorganisation energy of 36~meV. The lower the Drude-Lorentz peak frequency, the more networks meet this criterion.}
    \label{fig:ENAQT reorg double peak bar chart}
\end{figure}

The goal of this work has been to demonstrate the existence of a new phenomenon in otherwise very typical transport networks. But answering this question is separate from demonstrating if this occurs at typical amounts of coupling to the environment. Also, answering this question cannot be done for the pure dephasing approach used in many studies, and instead requires microscopic couplings to be considered. As such we consider our Bloch-Redfield results in this section, and we calculate the reorganisation energies of our networks in the standard fashion

\begin{equation}
    \lambda = \frac{1}{\pi} \int^\infty _0 \frac{\mathcal{J}(\omega)}{\omega} d\omega,
\end{equation}
where $\lambda$ is the reorganisation energy and the other terms keep their definitions from \cref{eqn: noise-power-spectrum}. With this, we can constrain our results to only those occurring at physically reasonable reorganisation energies, which we take to be anything below 36~meV (290.4~cm$^{-1}$), covering previously calculated reorganisation energies for FMO~\cite{Saito2019Site-DependentProtein}. \Cref{fig:ENAQT reorg double peak bar chart} shows what percentage of networks with multiple transport efficiency maxima have both peaks below this upper bound.

\section{Efficiency of single and double peaked networks}
\label{sec: single vs double peak efficiency}

\begin{figure}[H]
    \centering
    \includegraphics[width = \linewidth]{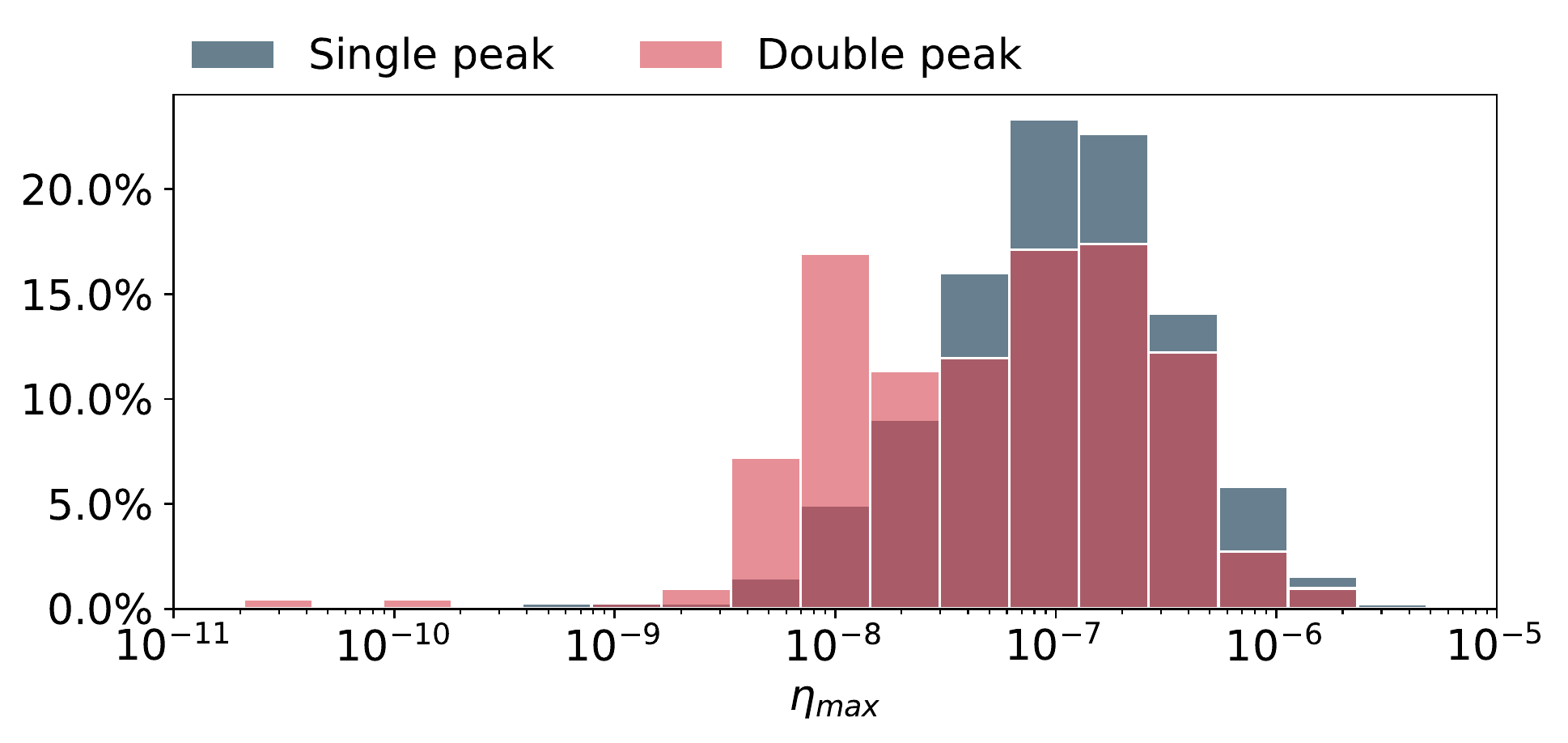}
    \caption{Bar chart showing the maximum efficiency $\eta_{max}$ of networks with either a single peak or multiple peaks in their transport efficiency. Y-axis shows the percentage of each dataset contained within each interval. We see a very strong overlap in the maximal efficiency of either network kind. While double peaked systems can have lower efficiencies, they also have a much larger spread of efficiencies than singly peaked systems.}
    \label{fig: single vs double peak efficiency}
\end{figure}

Here we consider the maximal efficiency $\eta_{max}$ of the networks in our main data with either a single peak or two peaks. As shown in \cref{fig: single vs double peak efficiency} we see a very large overlap in the efficiencies shown by either kind of transport efficiency landscape. Suggesting that double-peaked systems can be less efficient, but have such a spread of efficiencies that they are often equally as efficient as the singly peaked systems.

\section{Relative efficiency of each peak}
\label{sec: relative peak efficiency}
\begin{figure}[H]
    \centering
    \includegraphics[width = \linewidth]{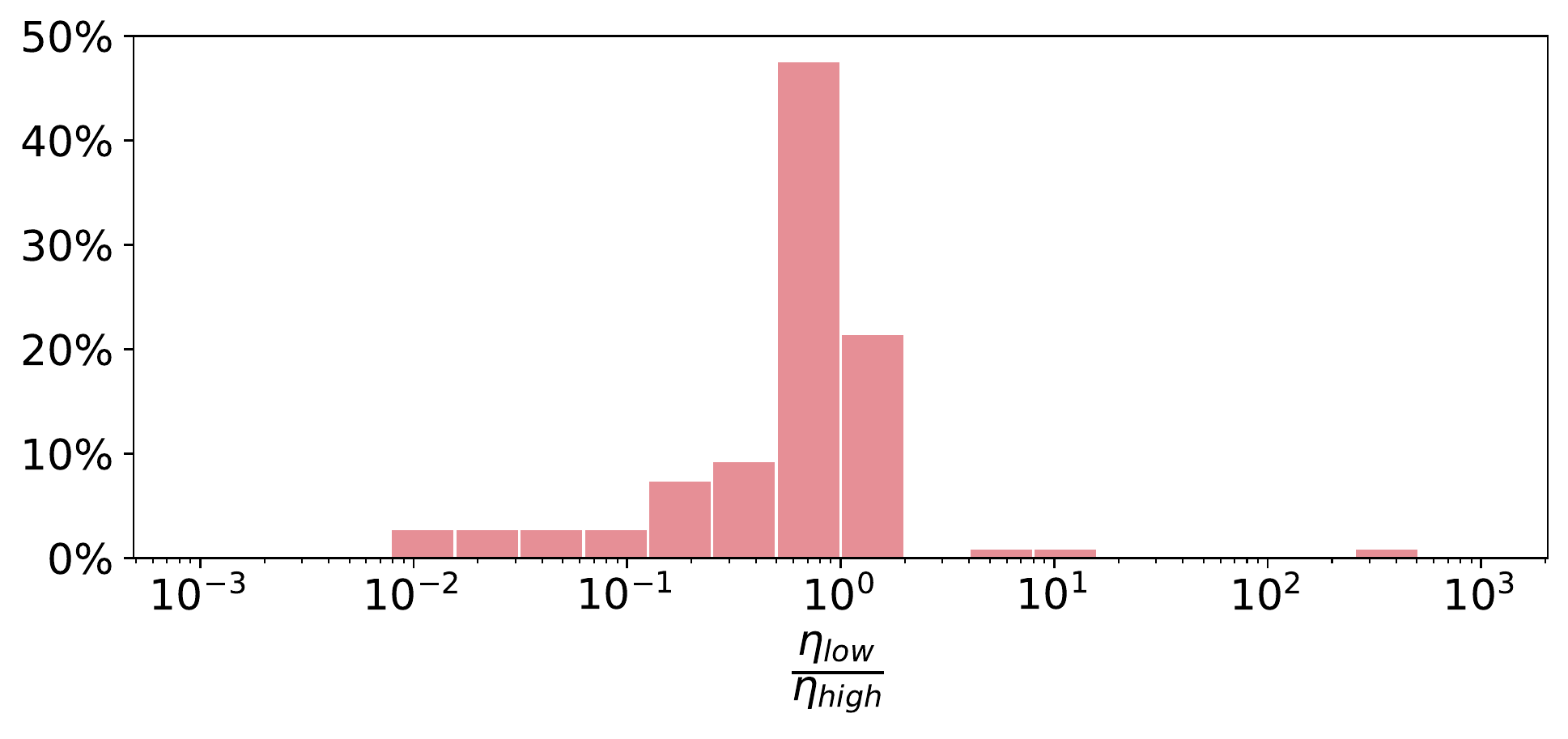}
    \caption{Histogram of the relative efficiency of systems with multiple ideal noise rates. Peaks at lower noise rates typically have less efficiency than those at higher noise rate. The central two bars contain the majority of systems and correspond to the two peaks having efficiencies within a factor of 2 of each other.}
    \label{fig: peak-efficiency-comparison}
\end{figure}

We now study the relative efficiency of both peaks in the double peaked case. We define the steady state transport efficiency of the low noise rate and high noise rate peak as $\eta_{low}$ and $\eta_{high}$ respectively. In \cref{fig: peak-efficiency-comparison} we consider the ratio of these two quantities $\frac{\eta_{low}}{\eta_{high}}$ for our main ensemble of energeticaly disordered networks.  

As \cref{fig: peak-efficiency-comparison} shows, the peaks at higher noise rates are typically more efficient than those at lower noise rates. However we also note that for the vast majority of systems the two peaks have efficiencies less than a factor 2 apart. The central two bars of the histogram correspond to the range $0.5 < \frac{\eta_{low}}{\eta_{high}} < 2$ and make up 69.1\% of the doubly peaked systems. So in most cases, both peaks have a similar prominence.

\section{Dense Networks}
\label{sec: dense-network}
In this work we have considered relatively sparse networks in a volume with a 10~nm radius. This meant there was lots of freedom for the dipoles and their respective exclusion volumes to form many different kinds of structures, while respecting the ideal dipole approximation that we used. However if one wants to understand light-harvesting complexes, these typically occur at higher densities.

\begin{figure}[H]
    \centering
    \includegraphics[width = \linewidth]{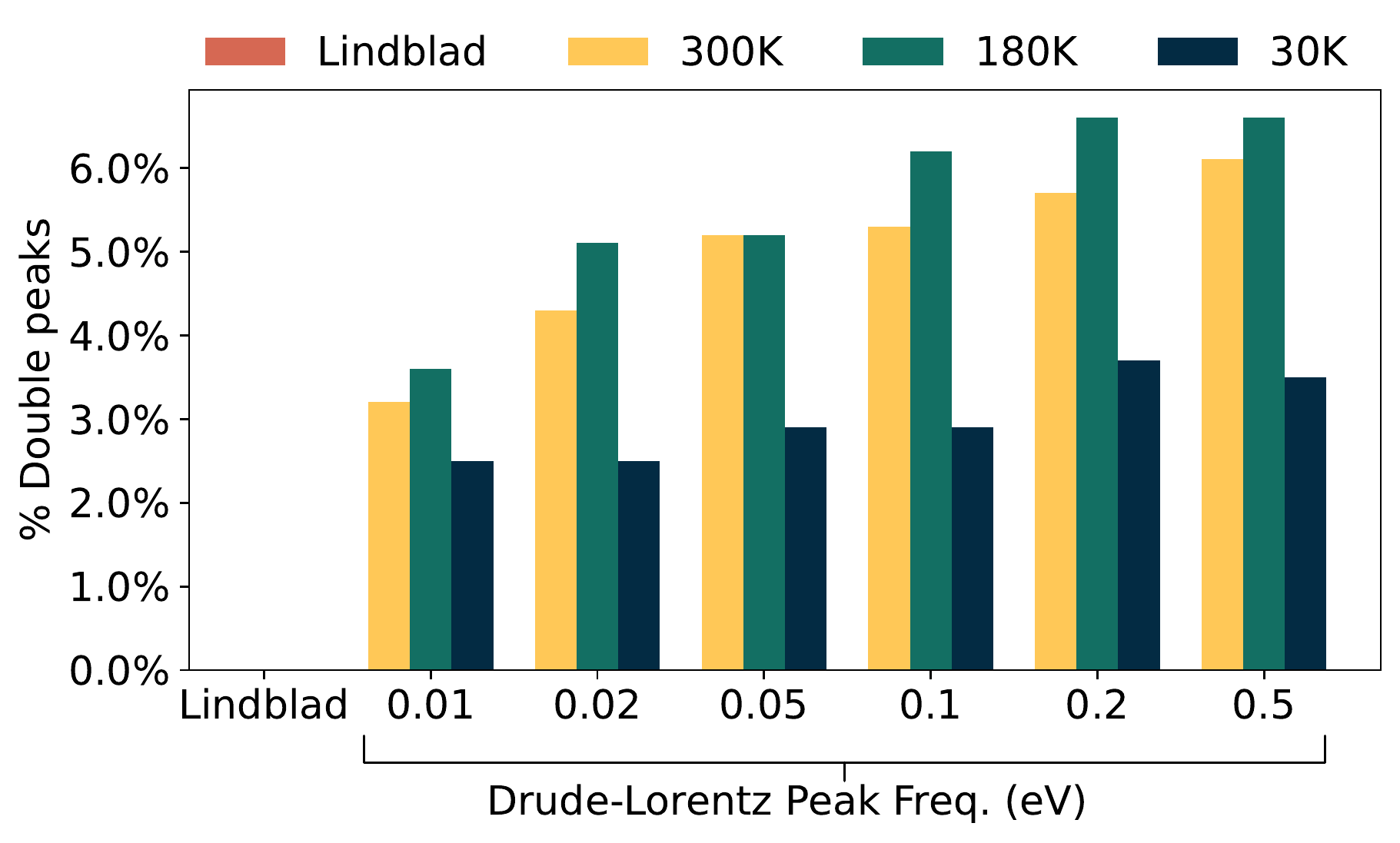}
    \caption{Double-peak rates for 1000 dense networks with a volume radius of 2.5nm and exclusion volume radius of 0.5nm. We observe no Lindblad peaks in these networks and see the highest rates of double peaks for intermediate temperatures of 180K.}
    \label{fig: dense-chromophore-bar}
\end{figure}

Here we briefly consider such a dense system, keeping the same model as before with 8 sites, but reduce the full sphere radius to 2.5 nm, close to prior work~\cite{Knee2017Structure-DynamicsSystems} and similarly reduce the exclusion volume around each site to 0.5 nm so that all 6 interior sites can still fit in the volume. The results are shown in \cref{fig: dense-chromophore-bar}, where we see comparable results to those in sparse networks, but with a decrease in the incidence of double peaks at low temperatures. 

\begin{figure}[H]
    \centering
    \includegraphics[width = \linewidth]{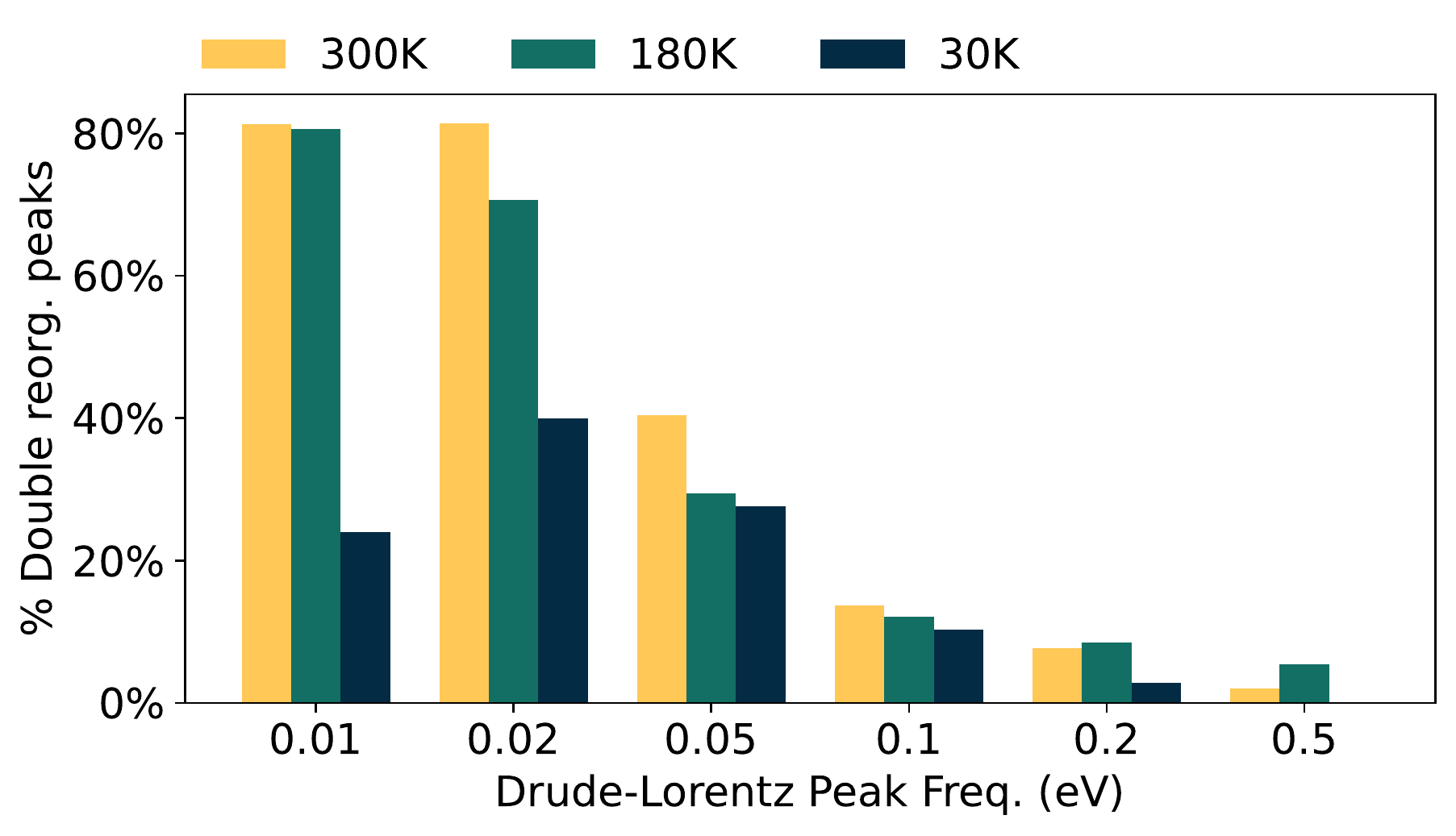}
    \caption{The percentage of double-peaked dense networks, where both peaks occur below the cutoff reorganisation energy of 36 meV.}
    \label{fig: dense-chromophore-reorg}
\end{figure}

The reorganisation energies of these peaks were also considered in \cref{fig: dense-chromophore-reorg}, where we see that the results at 30K are consistently the least likely to occur below the cutoff value.

\section{Analysing Data}
\label{sec: filtering}
In this work we have used both the Lindblad and non-secular Redfield master equations. By its mathematical construction the Lindblad master equation always produces completely positive density matrices which preserve the trace. Redfield master equations are also linearly trace preserving but can produce unphysical density matrices with negative probabilities~\cite{Jeske2015Bloch-RedfieldComplexes,Eastham2016Bath-inducedApproximation}. As this work is looking for peaks in arrays of values, we need to screen these erroneous points as sudden spikes or dips on the otherwise smoothly varying data produce false peaks.

To ensure the steady state populations were physical we checked if the trace was unitary and if all the on-site populations were positive. To ensure the steady states were valid we recorded the eigenvalues of each Redfield tensor steady state and then checked their eigenvalues were all between 0 and 1. Tolerances of 10$^{-5}$ were used for the site checks, and 10$^{-4}$ for the eigenvalue checks as these were sufficient to remove erroneous points. The points excluded occurred at higher system-environment couplings, while results at lower couplings were rarely if ever excluded. These checks were also applied to the Lindblad results for consistency. With these points removed, simple peak finding algorithms were applied to the remaining valid points in each array: no requirements were placed on the peak prominence, heights or widths.

\end{document}